\documentclass[journal,twocolumn,letterpaper]{IEEEJERM}

\usepackage{xcolor}
\usepackage{balance}
\usepackage[cmex10]{amsmath}

\usepackage{amsmath}
\usepackage{bm}
\usepackage{empheq}
\usepackage{amssymb}
\usepackage{multirow}
\usepackage{ifpdf}
\usepackage{stfloats}
\usepackage{enumitem}
\usepackage{textcomp}

\usepackage{algorithmic}
\usepackage{array}
\usepackage{mdwmath}
\usepackage{mdwtab}
\usepackage{eqparbox}
\usepackage{cite} 
\usepackage{comment}
\usepackage{url}
\usepackage[pdftex]{graphicx}
\graphicspath{{immagini/}}  

\usepackage{tikz}

\usepackage{stfloats}
\usepackage{url}

\newcommand{\ped}[1]{_\textsc{\scriptsize{#1}}}
\newcommand{\ap}[1]{^\textsc{\scriptsize{#1}}}
\newcommand{\timeT}{\mathcal{T}}

\newcommand\copyrighttext{%
  \footnotesize \textcopyright 2020 IEEE. Personal use of this material is permitted. Permission from IEEE must be obtained for all other uses, in any current or future media, including reprinting/republishing this material for advertising or promotional purposes, creating new collective works, for resale or redistribution to servers or lists, or reuse of any copyrighted component of this work in other works.}
\newcommand\copyrightnotice{%
\begin{tikzpicture}[remember picture,overlay]
\node[anchor=south,yshift=3pt] at (current page.south) {\fbox{\parbox{\dimexpr\textwidth-\fboxsep-\fboxrule\relax}{\copyrighttext}}};
\end{tikzpicture}%
}

\begin{document}
	
	\title{Fast Optimization of Temperature Focusing in
		Hyperthermia Treatment of Sub-Superficial Tumors}
	
	\author{Rossella~Gaffoglio,
		Marco~Righero,
		Giorgio~Giordanengo,
		Marcello~Zucchi,
		and~Giuseppe~Vecchi,~\IEEEmembership{Fellow,~IEEE}%
	}
	
	\markboth{IEEE Journal of Electromagnetics, RF and Microwaves in Medicine and Biology}
	{Gaffoglio \MakeLowercase{\textit{et al.}}: Temperature Focusing in Microwave Cancer Hyperthermia via Pre-Corrected SAR-Based Focusing}
	
	\twocolumn[
	\begin{@twocolumnfalse}
		
		\maketitle
		
\begin{abstract}
Microwave hyperthermia aims at selectively heating cancer cells to a supra-physiological temperature. For non-superficial tumors, this can be achieved by means of an antenna array equipped with a proper cooling system (the water bolus) to avoid overheating of the skin. In patient-specific treatment planning, antenna feedings are optimized to maximize the specific absorption rate (SAR) inside the tumor, or to directly maximize the temperature there, involving a higher numerical cost. 
We present here a method to effect a low-complexity temperature-based planning. It arises from recognizing that SAR and temperature have shifted peaks due to thermal boundary conditions at the water bolus and for physiological effects like air flow in respiratory ducts. In our method, temperature focusing on the tumor is achieved via a SAR-based optimization of the antenna excitations, but optimizing its target to account for the cooling effects. The temperature optimization process is turned into finding a SAR peak position that maximizes the chosen temperature objective function. Application of this method to the 3D head and neck region provides a temperature coverage that is consistently better than that obtained with SAR-optimization alone, also considering uncertainties in thermal parameters. This improvement is obtained by solving the bioheat equation a reduced number of times, avoiding its inclusion in a global optimization process. 
\end{abstract}

		
\end{@twocolumnfalse}]
	
	{
		\renewcommand{\thefootnote}{}%
		\footnotetext[1]{This manuscript was submitted for review on May 25, 2020; revised on July 18, 2020 and September 10, 2020. This work was supported by the MIUR PRIN 2015KJE87K. Preliminary findings of this work were submitted to the 14th European Conference on Antennas and Propagation (EuCAP), 2020.}
		\footnotetext[2]{R. Gaffoglio, M. Righero and G. Giordanengo are with the Advanced Computing and Applications area, Fondazione LINKS, 10138 Turin, Italy (e-mails: rossella.gaffoglio@linksfoundation.com, marco.righero@\-linksfoundation.com, giorgio.giordanengo@linksfoundation.com).}
		\footnotetext[3]{M. Zucchi and G. Vecchi are with the Department of Electronics and Telecommunications, Politecnico di Torino, 10129 Turin, Italy (e-mails: marcello.zucchi@polito.it, giuseppe.vecchi@polito.it).}
	}

	\copyrightnotice

	\IEEEpeerreviewmaketitle

	\section{Introduction}
	\IEEEPARstart{M}{icrowave} hyperthermia therapy is a selective increase of the tumor cells temperature to 42-43$^\circ$C induced by a proper antenna system; 
	it is recognized as a potent sensitizer for current cancer treatments \cite{Kampinga,Hurwitz2014}. As demonstrated by several clinical trials \cite{Cihoric,Datta,Datta2016neck,Datta2016cervix,Bakker,Kroesen}, the targeted increase of the tumor temperature leads to an improvement in the clinical effectiveness of radiotherapy and chemotherapy, with no significant acute or late additional morbidities. 
	
	In the state of the art for internal tumors, heating of the target is achieved by exploiting constructive wave interference from a certain number of antennas \cite{PaulidesHYP,Crezee,Drizdal}. Wave physics implies that heating of the skin is inevitable; to avoid this, a bag filled with circulating demineralized water (the so called water bolus) is introduced between the applicator and the body \cite{Gaag,Rijnen}.
	
	The effectiveness of a hyperthermia treatment is strongly related to the possibility to plan and control the administrated heating \cite{Kok_2015}, especially for challenging anatomical sites such as the head and neck (H\&N) region \cite{Paulides_2016}. Treatment planning is thus based on patient-specific numerical simulations \cite[e.g.]{Paulides_2013}, employing  3D CT or MRI scans, together with the applicator model. The optimization of the antenna feedings is performed using SAR-based \cite{Rijnen_april_2013,Iero,Bellizzi} or temperature-based \cite{Kok_2005} (T-based) optimization methods \cite{Kok_2015}. 
	
	SAR-based optimization, as in \cite{Rijnen_april_2013,Iero,Bellizzi}, does not control temperature directly, but there is an assessed correlation between SAR indicators, temperature, and clinical outcome \cite{Lee_1998,Canters1}. It is to be expected that deviations between temperature (T) and SAR distributions are due to thermal boundary conditions, arising from external cooling (i.e. the water bolus) and physiology, like the air flow in respiratory tracts, and possibly major blood vessels. This consideration is key to the strategy presented here. 
	
	T-based techniques aim at directly optimizing the quantity of interest, i.e. temperature distribution, although their effectiveness may be reduced by the large uncertainty of some thermal tissue parameters \cite{Canters_2013,Greef}. This approach has a higher computational cost \cite{Kok_2015} than the SAR-based one, as at any step of the optimization one solves for the thermal equation; this has prompted the development of sophisticated numerical techniques to reduce the computational burden \cite{Das,Kok_2013}. 
	
	Recognizing thermal boundary conditions as the main source of the SAR-T deviation, we present a scheme to employ SAR-based techniques, but whose goal is optimized to achieve the desired T distribution in the presence of the relevant thermal boundary conditions. Preliminary results of this method were presented in \cite{EuCAP2020}. Here, the algorithm has been improved and rationalized, and its description accordingly recast; moreover, the key assumptions of the procedure have been verified.
	
	The paper is organized as follows. Section II summarizes heat transfer in tissues; Section III describes the implemented numerical testbed concerning a tumor in the H\&N region, the SAR focusing,	and the corresponding thermal analysis. The proposed procedure for temperature focusing based on SAR optimization is presented in Section IV, with results in Section V, and conclusions in Section VI.

	\section{Bioheat Transfer and Boundary Conditions} \label{sec_BHeq}
	
	\begin{figure}[!t]
		\begin{center}
			\includegraphics[width=0.48\textwidth]{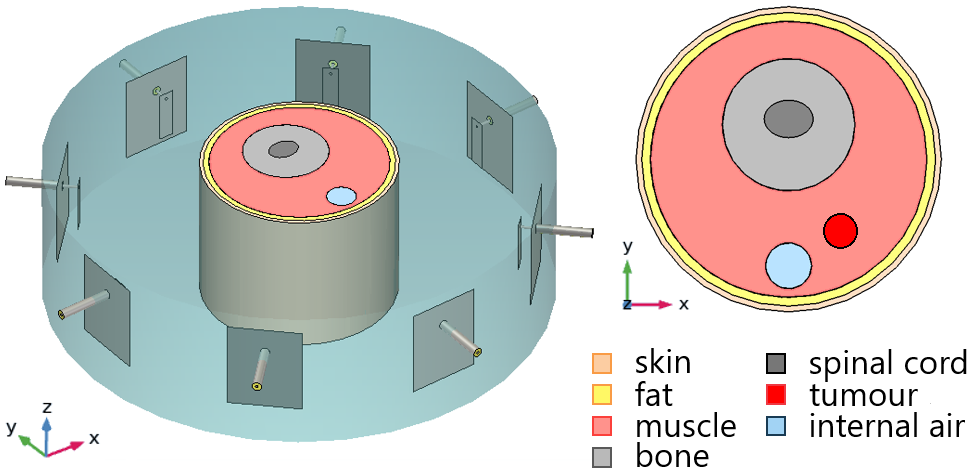}
			\caption{{\it Left}: geometry of the phased array applicator, the simple neck phantom and the surrounding water bolus; {\it right}: middle top view ($z=0$ plane) of the simplified neck model with all the considered tissues.}
			\label{fig_JERM-2020-05-0074_Fig1}
		\end{center}
	\end{figure}
	
	The temperature increase in a tissue caused by the exposure to an external heating source can be described as usual using Pennes' bioheat equation \cite{Pennes}; as common in the hyperthermia planning literature \cite{Paulides_2013}, we will be concerned with the steady-state version of the equation: 
	\begin{equation} \label{eq_fII1}
	-\bm{\nabla}\cdot(k(\bm{r})\:\bm{\nabla}T)=f(\bm{r}),
	\end{equation}
	where $T$ $\left(^\circ\mbox{C}\right)$ is the temperature, $k$ $\left(\mbox{W}/(\mbox{m}^\circ\mbox{C})\right)$ is the tissue-specific thermal conductivity, and $f(\bm{r})$ $\left(\mbox{W}/\mbox{m}^3\right)$ is the source term, given by:
	\begin{equation} \label{eq_fII3}
	f(\bm{r})=Q_{hs}(\bm{r})+\rho_b\:C_{p,b}\:\omega(\bm{r})\:(T(\bm{r})-T_a).
	\end{equation}
	In (\ref{eq_fII3}), $Q_{hs}$ $\left(\mbox{W}/\mbox{m}^3\right)$ is the external heat source, $\rho_b$ $\left(\mbox{kg}/\mbox{m}^3\right)$ and $C_{p,b}$ $\left(\mbox{J}/(\mbox{kg}^\circ\mbox{C})\right)$ are the blood density and specific heat, $T_a$ $\left(^\circ\mbox{C}\right)$ is the arterial blood temperature, and $\omega$ (1/s) is the volumetric blood perfusion rate, often alternatively expressed as $w=(\omega/\rho)\cdot 60\mbox{e}6$ (ml/(min kg)), being $\rho$ $\left(\mbox{kg}/\mbox{m}^3\right)$ the tissue-specific mass density. The metabolic heat generation term has been neglected in (\ref{eq_fII3}) as usual.
	In the presence of an electromagnetic source (such as an antenna system), the heat source $Q_{hs}$ can be written as:
	\begin{equation} \label{eq_fII4}
	Q_{hs}(\bm{r})=\frac{1}{2}\:\sigma(\bm{r})\:|\bm{E}(\bm{r})|^2=\rho(\bm{r})\:\mbox{SAR}(\bm{r}),
	\end{equation}
	where $\bm{E}$ (V/m) is the electric field (peak value), $\sigma$ (S/m) is the electrical conductivity, and SAR (W/kg) denotes the specific absorption rate. The values of dielectric and thermal properties are reported in literature as averages over various studies and measurements; therefore, they represent a source of uncertainty \cite{Balidemaj,Verhaart_nov_2014}. Perfusion, in particular, is more crucial in the estimation of temperature profiles, being characterized by the largest uncertainty \cite{Greef}.
	
	The thermal model has to be completed by the boundary conditions (B.C.) for the temperature $T$, specifying how the system interacts with the outside environment.
	
	For the realistic case of a finite region, two basic types of thermal boundary conditions can be identified: specified temperature (isothermal condition) and specified heat flux. The isothermal condition is appropriate when the body is in contact with a good heat conductor maintained at a constant temperature $T_s$ $\left(^\circ\mbox{C}\right)$ (such as an ice pad) \cite{Gao} and it is given by:
	\begin{equation} \label{eq_fII5}
	T(\bm{r}) = T_s, \ \ \ \bm{r}\in\partial\mathcal{V},
	\end{equation}
	where $\partial\mathcal{V}$ indicates the boundary of the considered volume region $\mathcal{V}$. 
	The effects of an exterior fluid cooling the surface of the body are instead better described by a convective heat flux boundary condition:
	\begin{equation} \label{eq_fII6}
	\bm{\hat{n}}\cdot(k(\bm{r})\:\bm{\nabla}T(\bm{r})) = h\:(T_s-T(\bm{r})), \ \ \ \bm{r}\in\partial\mathcal{V},
	\end{equation}
	where $\bm{\hat{n}}$ is the unit vector normal to the boundary, $k$ $\left(\mbox{W}/(\mbox{m}^\circ\mbox{C})\right)$ is the local thermal conductivity, $T_s$ $\left(^\circ\mbox{C}\right)$ is the external temperature and $h$ $\left(\mbox{W}/({\mbox{m}^2}\right.$$\left.^{\circ}\mbox{C})\right)$ is the heat transfer coefficient.

	\section{SAR focusing and temperature shift} \label{sec_SARTemp}
	
	\begin{table}[!t]
		\renewcommand{\arraystretch}{1.3}
		\caption{Thermal and dielectric properties at $f=434$ MHz \cite{Hasgall}}
		\label{tab_param}
		\centering
		\begin{tabular}{l|c|c|c|c|c}
			\hline
			\hline
			\multirow{2}{*}{\textbf{Tissue}} & $\rho$ & $k$ & $w$ & $\epsilon_r$ & $\sigma$ \\
			& $\left[\frac{\mbox{kg}}{\mbox{m}^3}\right]$ & $\left[\frac{\mbox{W}}{\mbox{m}^{^\circ}\mbox{C}}\right]$ & $\left[\frac{\mbox{ml}}{\mbox{min kg}}\right]$ & [-] & [S/m] \\ [1ex]
			\hline
			\hline
			Skin (wet)                              & 1109 & 0.37  & 106   & 49.4\textsuperscript{\cite{Andreuccetti}} & 0.681\textsuperscript{\cite{Andreuccetti}} \\
			Fat                                     &  911 & 0.21  &  33   & 11.6 & 0.082 \\
			Muscle\textsuperscript{\cite{Drizdal}}  & 1090 & 0.49  &  39.1 & 56.9 & 0.805 \\
			Bone                                    & 1908 & 0.32  &  10   & 13.1 & 0.094 \\
			Spinal cord                             & 1075 & 0.51  & 160   & 35   & 0.456 \\
			Tumor\textsuperscript{\cite{Drizdal}}   & 1050 & 0.51  &  72.3 & 59   & 0.89  \\
			Internal air                            & 1.15 & 0.026 &   0   &  1   & 0     \\
			\hline
			\hline
		\end{tabular}
	\end{table}
	
	\subsection{Numerical testbed} \label{sec_model}
	
	\begin{figure}[!b]
		\centering
		\includegraphics[width=3.45in]{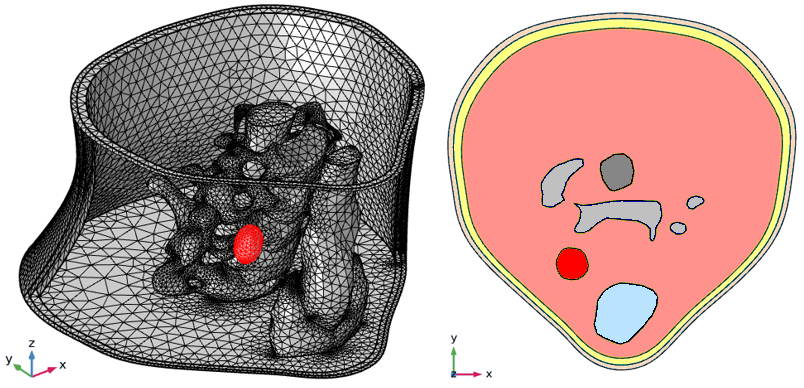}
		\caption{{\it Left}: realistic neck phantom; {\it right}: middle top view ($z=0$ plane) of the neck model with all the considered tissues (see the legend of Fig. \ref{fig_JERM-2020-05-0074_Fig1}).}
		\label{fig_JERM-2020-05-0074_Fig2}
	\end{figure}
	
	We implemented a numerical model to study the application of the microwave heating to a tumor placed in the human neck, using the finite-element solver COMSOL Multiphysics \cite{Comsol} for both the solution of the Electromagnetic and Bioheat problems; it ran on an Intel\textsuperscript\textregistered $\,$ Core\textsuperscript\texttrademark $\,$ i7-7700 workstation, with 64 GB RAM. All code has been written in MATLAB, without optimization.
	
	For the sake of illustrating issues and the proposed method, we will first consider a simplified computational phantom of the human neck, where the vertebrae, the trachea, the spinal cord, as well as the neck shape are modeled by means of cylinders, while the tumor is represented by a sphere of diameter $d=12$ mm; this is shown in Fig. \ref{fig_JERM-2020-05-0074_Fig1}. Then, our method will be tested on the more realistic neck computational model visualized in Fig. \ref{fig_JERM-2020-05-0074_Fig2}. This phantom has been realized using the 3D mesh models of the vertebrae (from C3 to C7), the spinal cord and the laryngotracheal canal provided by the Visible Human Project (VHP Female Version 2.2 \cite{Makarov}). In this case, the GTV has been simulated as a mass with irregular shape, being $d=10$ mm the diameter of the smallest enclosing sphere.
	
	A phased array applicator is a common choice for the heating of H\&N tumors \cite{Crezee,Drizdal}. With reference to \cite{PaulidesHYP}, we considered a uniform circular array made of $N=8$ patch antennas with water substrate, immersed in a water bolus and properly matched to operate at the frequency $f=434$ MHz \cite{PaulidesPatch} at a distance of 5 cm from the neck phantom. The length and the width of the optimized patch antennas are 27.4 mm and 9.8 mm, respectively, the height is 7 mm, and the distance of the feed to the edge is 4.5 mm. 
	
	The dielectric and thermal properties were assigned to the phantom according to the material database in \cite{Hasgall}, using a tumor perfusion value 1.85 times higher than the rest value for the muscle perfusion \cite{Lang} (see Table \ref{tab_param}).
	
	\begin{figure*}[!t]
		\begin{center}
			\includegraphics[width=1\textwidth]{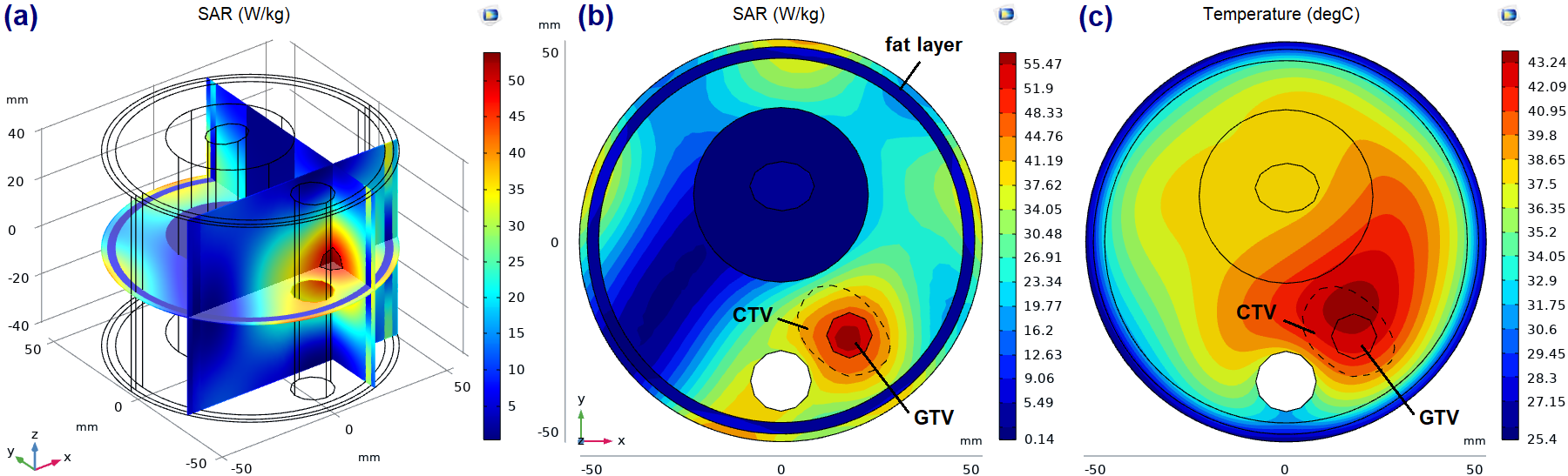}
			\caption{SAR-Temperature shift. Panel (a): SAR distribution optimized on the GTV. Panel (b): SAR optimized on the GTV displayed on the $z=0$ plane; the clinical target volume (CTV) is also reported. The blue annulus corresponds to the fat layer, characterized by a small power deposition due to the low electrical conductivity $\sigma$ (see Table \ref{tab_param}). Panel (c): temperature distribution on the $z=0$ plane, achieved solving the bioheat equation on the whole neck region with a convective flux boundary condition on the tissue-water bolus interface and the SAR profile of panel (a) as source term.}
			\label{fig_JERM-2020-05-0074_Fig3}
		\end{center}
	\end{figure*}
	
	The model boundary conditions were set as per the current literature \cite{Verhaart_nov_2014}. A convective heat flux boundary condition (\ref{eq_fII6}) was applied on the tissue-water bolus interface, with $T_{s}=20\:^\circ$C and $h=82$ $\mbox{W}/({\mbox{m}^2}$$\:^{\circ}\mbox{C})$, and on the internal boundaries of the laryngotracheal canal, with $T_{s}=30\:^\circ$C \cite{McFadden} and $h=50$ $\mbox{W}/({\mbox{m}^2}$$\:^{\circ}\mbox{C})$. Moreover, the isothermal condition $T(\bm{r})=37\:^\circ$C was imposed on the upper and lower surfaces of the neck, being reasonable to assume that the temperature of the head and the body trunk remains approximately constant during treatment in real applications.

	\subsection{SAR-based optimization} \label{sec_SAR}
	In this paper, we consider the SAR-based optimization approach adopted in \cite{Rijnen_april_2013}. 
	The total electric field generated by the array is written as a superposition of unknown excitation coefficients and the electric fields generated by each antenna acting as standalone. In this way, one can solve initially for all single-feed antenna fields, and let the global optimization only handle linear combinations of these fields (see e.g. \cite{Rijnen_april_2013}). 
	
	The implemented optimization aims at finding the excitation coefficients that maximize the SAR in the gross target volume (GTV), i.e., the tumor, minimizing the risk of hotspots in the surrounding healthy tissues. The corresponding cost function is the target-to-hotspot SAR quotient (THQ) \cite{Rijnen}:
	\begin{equation} \label{eq_fIII1}
	\mbox{THQ}=\frac{\left\langle \mbox{SAR}\ped{\:target}\right\rangle}{\left\langle \mbox{SAR}\ped{\:V1}\right\rangle},
	\end{equation}
	where $\left\langle\mbox{SAR}\ped{\:target}\right\rangle$ is the average SAR in the target region (GTV) and $\left\langle \mbox{SAR}\ped{\:V1}\right\rangle$ is the average SAR in V1, which is the $1\%$ of the healthy volume with the highest SAR (hotspot SAR) \cite{Rijnen,Canters1}. We minimized the function 1/THQ with the Particle Swarm Optimization (PSO) algorithm provided by MATLAB \cite{Matlab}.  
	
	For the simplified model of Section \ref{sec_model}, the procedure has led to a good SAR focusing on the gross target volume (GTV) represented by the sphere, as shown in Figs. \ref{fig_JERM-2020-05-0074_Fig3}.a, \ref{fig_JERM-2020-05-0074_Fig3}.b, where the optimized SAR distribution is reported.

	\subsection{SAR-Temperature shift} \label{sec_shift}
	The SAR obtained as above was then inserted as source in the bioheat equation \eqref{eq_fII1}. The resulting temperature map is reported in Fig. \ref{fig_JERM-2020-05-0074_Fig3}.c. As can be observed, a significant temperature shift outside the GTV occurs. This shift can be attributed to the combined effect of the thermal boundary conditions dictated by the water bolus cooling and by the air flux in the trachea; indeed, we have ascertained that by removing the air flux the shift occurs in the direction normal to the skin (with water bolus).
	
	\begin{figure*}[!t]
		\centering
		\includegraphics[scale=0.54]{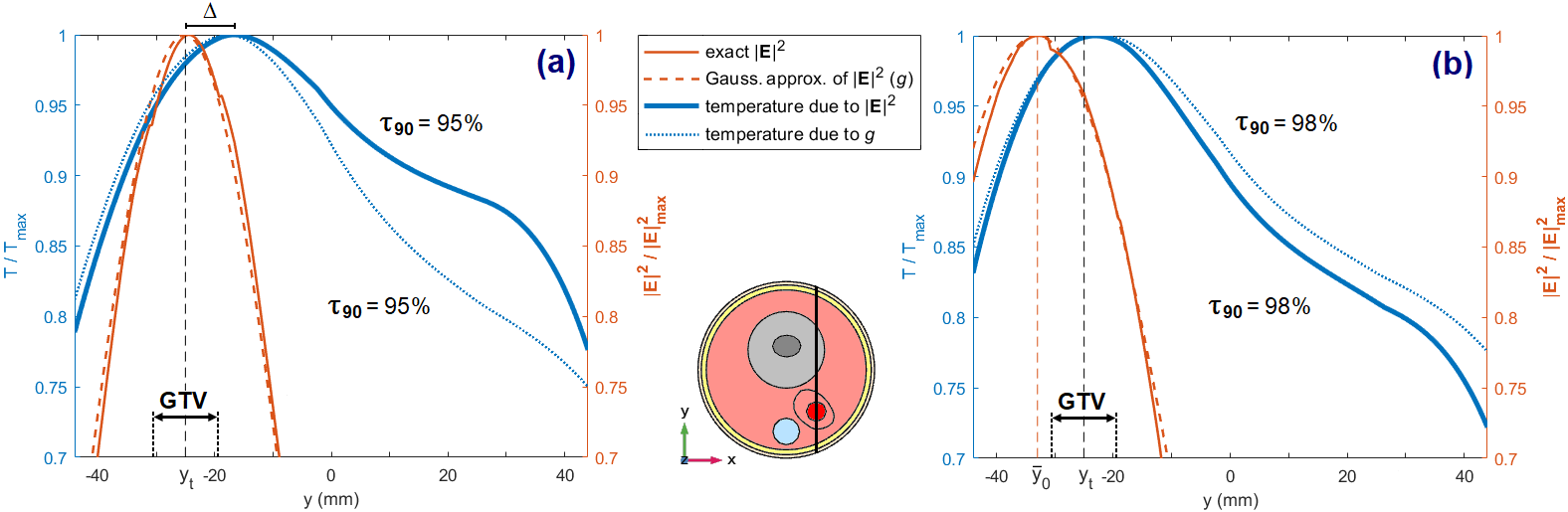}
		\caption{Gaussian approximation for SAR. Exact squared amplitude of the electric field and its Gaussian approximation, with the corresponding temperature distributions, evaluated along the line connecting the peaks of SAR and temperature (here along the $y$-axis) and reported for: (a) SAR optimized on the GTV; (b) SAR optimized for focusing on a different location. The extension of the GTV along the $y$-axis is highlighted.}  
		\label{fig_JERM-2020-05-0074_Fig4}
	\end{figure*}

	\section{T-shaping via SAR focusing: Methodology} \label{sec_Correction}

	\subsection{Objective function for T-based optimization}
	We begin by defining a goal (objective) function for our T-based optimization of antenna excitations. In practical applications, a good shaping involves both temperature ``focusing" (i.e., maximizing T in the desired GTV) and temperature uniformity; this is achieved by optimizing a single parameter: 
	\begin{equation} \label{eq_fIV3}
	\tau_{90}=\frac{\mbox{T}90}{\mbox{max}_{\bm{r}\in{\textsc{\scriptsize{GTV}}}}\:\left\{T(\bm{r})\right\}}.
	\end{equation}
	where the T90 parameter \cite{Verhaart_june_2014,Canters1} is defined as the temperature exceed by 90\% of the points in the GTV. 
	Expression (\ref{eq_fIV3}) will be our objective function.

	\subsection{Strategy} \label{sec_strategy}
	The analysis of the SAR-T shift due to thermal boundary conditions (Section \ref{sec_shift}) provides the ground to frame our simplified T-based optimization method. We view T-based planning as a modified SAR-based planning, in which the SAR goal is defined to achieve the desired T goal.  
	The steps of the proposed strategy can be summarized as follows:
	\begin{enumerate}[leftmargin=*,align=left]
		\item Numerical evaluation of the fields generated by the array feeding one antenna at a time (Maxwell's equations). 
		\item SAR-based optimization that maximizes the SAR in the GTV (Section \ref{sec_SAR}).
		\item Approximation of the resulting E-field squared magnitude with a (multi-variate) Gaussian function, used to compute a Gaussian heat source (\ref{eq_fII4}), with variable peak position $\bm{r}_0$.
		\item Definition and discretization of a refinement region $\mathcal{V}_\mathcal{R}$, surrounding the GTV.
		\item Solution of the bioheat equation (\ref{eq_fII1}) and evaluation of the objective function $\tau_{90}$ (\ref{eq_fIV3}) for each point in $\mathcal{V}_\mathcal{R}$, i.e., when the center of the Gaussian heat source (SAR mask) is moved inside the refinement region.
		\item Selection of the point $\bm{\bar{r}}_0$ in $\mathcal{V}_\mathcal{R}$ corresponding to the highest value of the objective function $\tau_{90}$.
		\item SAR-based optimization that maximizes the SAR in a target sphere centered around the optimized position $\bm{\bar{r}}_0$.
	\end{enumerate}
	
	The numerical complexity of the proposed approach is discussed in Sec.~\ref{sec_NumCompl}.
	We now describe the main steps of the above-reported procedure in more detail.

	\subsection{Gaussian fitting} \label{sec_fitting}
	As alluded above, we recognized by inspection that the focused SAR distribution can be well approximated by a multivariate Gaussian profile in the region of interest. More specifically, the Gaussian approximation is done on the squared amplitude of the electric field and not on the SAR itself. 
		
	Hence, we fit the squared amplitude of the electric field, $|\bm{E}(\bm{r})|^2$, with a multivariate Gaussian function with height $a$, center $\bm{r}_0$, and width $\Sigma=\mbox{diag}(\sigma_x,\sigma_y,\sigma_z)$, being $\sigma_x$, $\sigma_y$, $\sigma_z$ the standard deviations along the different axes:
	\begin{equation} \label{eq_fIV5}
	g(\bm{r};\bm{r}_0,a,\Sigma)=a\:\exp\left[-\frac{1}{2}\:(\bm{r}-\bm{r}_0)^{\textsc{\tiny{T}}}\:\Sigma^{-2}\:(\bm{r}-\bm{r}_0)\right].
	\end{equation}
	This will allow to insert a source term in the bioheat equation \eqref{eq_fII1} to perform the optimization of the SAR focusing position.
	
	The suitability of this local approximation is readily assessed by the temperature goal function $\tau_{90}$, which is equal to 95\% for both the actual SAR distribution and its Gaussian approximation (Fig. \ref{fig_JERM-2020-05-0074_Fig4}.a).
	
	Although being an approximation, different numerical tests showed us that the Gaussian function fits well an optimized SAR distribution in the region of interest. However, the proposed method does not prevent from considering other fitting functions, if better suited.

	\subsection{Optimum SAR focusing point search}
	Thanks to the Gaussian approximation of the electric field squared amplitude, using (\ref{eq_fIV5}), the center of the Gaussian SAR function can be easily moved within a region $\mathcal{V}_{\mathcal{R}}$ surrounding the GTV and the corresponding temperature map can be calculated accordingly, solving the bioheat equation (\ref{eq_fII1}); for each position of the SAR center, one thus computes the temperature distribution, and from this the $\tau_{90}$ objective function (\ref{eq_fIV3}). From the list of these values one picks the SAR peak location which yields the largest value of the objective function. This requires to solve the thermal equation for a very reduced number of times (grid of points in the refinement region), in the context of a convex search of the optimum point.
	
	In this exhaustive search, it is necessary to establish a) the extension of the search region $\mathcal{V}_{\mathcal{R}}$, called refinement region, and b) the sampling density in this region. The distance $\Delta$ between the SAR and temperature peaks appears as a meaningful characteristic length of our problem.
	
	It is also reasonable to consider a refinement region $\mathcal{V}_{\mathcal{R}}$ with the shape of a ball centered around the tumor centroid $\bm{r}_t$. We observe that a refinement region much larger than $\Delta$ could affect the significance of the $\tau_{90}$ parameter; the diameter of $\mathcal{V}_{\mathcal{R}}$ is suggested to be: $d_{\scriptstyle{\mathcal{R}}}=d+2\Delta$, being $d$ the radius of the smallest sphere enclosing the GTV. The volume of $\mathcal{V}_{\mathcal{R}}$ can be discretized with a sampling distance $\Delta/\mathcal{N}$ among points, where $\mathcal{N}$ indicates the sampling density. 
	
	\begin{figure*}[!t]
		\centering
		\includegraphics[width=1\textwidth]{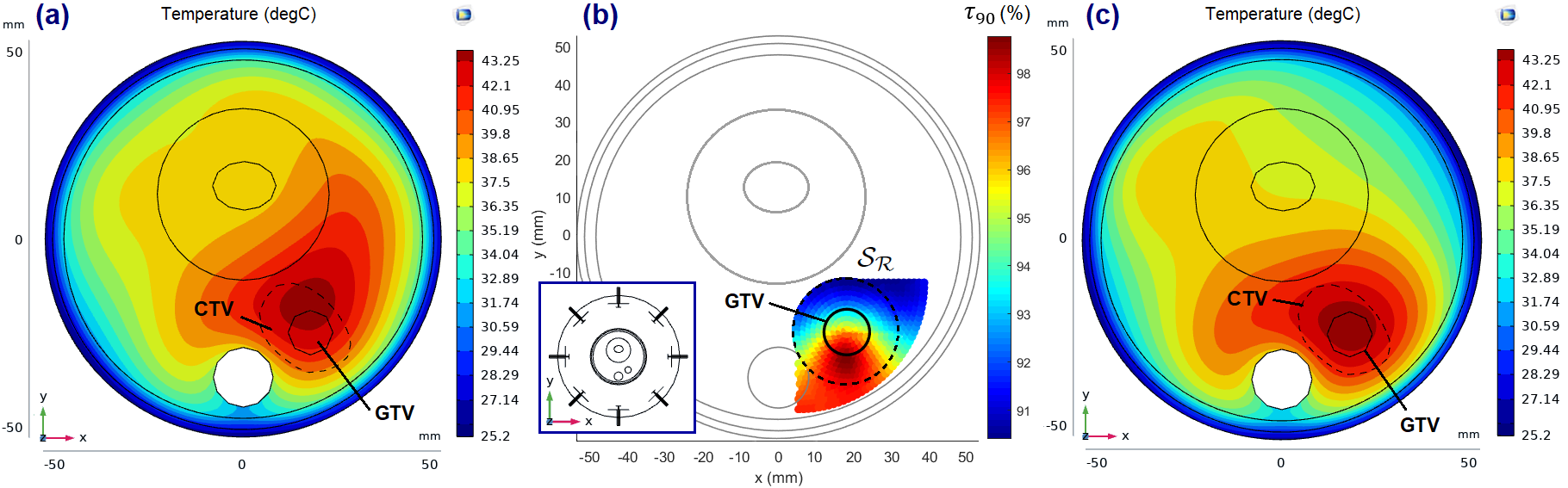}
		\caption{Effect of optimization of SAR target position. Panel (a): temperature distribution on the $z=0$ plane before the application of the corrective procedure. Panel (b): objective function $\tau_{90}$ as a function of the Gaussian SAR focusing center for the two-dimensional refinement region $\mathcal{S}_\mathcal{R}$ (dashed circle) - a larger region is shown for inspection. The objective function value is represented in color scale. The highlighted small circle inside the refinement region indicates the tumor circular section (GTV) on the $z=0$ plane. Panel (c): temperature distribution on the $z=0$ plane for the SAR map obtained by optimizing the antenna weights to focus the SAR in the region centered at the pre-corrected location $\bar{\bm{r}}_0=(18,-33,0)$ mm.}
		\label{fig_JERM-2020-05-0074_Fig5}
	\end{figure*}
	
	Using an exhaustive search approach, the SAR map:
	\begin{equation} \label{eq_SARmap}
	f_{\textsc{\tiny{SAR}}}(\bm{r};\bm{r}_0,a,\Sigma)=\frac{1}{2\rho(\bm{r})}\:\sigma(\bm{r})\:g(\bm{r};\bm{r}_0,a,\Sigma),
	\end{equation}
	with $a$ and $\Sigma$ fixed by the Gaussian fit described in Section \ref{sec_fitting}, is generated for each point $\bm{r}_0\in\mathcal{V}_\mathcal{R}$, i.e., the Gaussian function center is moved on the grid of sampling points inside $\mathcal{V}_\mathcal{R}$, the bioheat equation \eqref{eq_fII1} is solved, and the $\tau_{90}$ parameter is evaluated. The effect of this search, here limited to a two-dimensional ball surrounding the GTV on the $z=0$ plane, is shown in Fig. \ref{fig_JERM-2020-05-0074_Fig5}.b.

	\section{Results and Discussion} \label{sec_Results}
	
	\subsection{Simple neck model} \label{sec_Simple}
	For the case under study, we test the proposed strategy limiting the refinement region to a two-dimensional ball $\mathcal{S}_\mathcal{R}$ on the plane of the SAR-temperature shift; with our coordinate system this is on the $z=0$ plane and centered at $\bm{r}_t = (18,-25,0)$ mm, with a diameter $d_{\scriptstyle{\mathcal{R}}}=d+2\Delta=2.8$ cm, being $d=12$ mm the tumor sphere diameter and $\Delta=8$ mm the SAR-temperature shift magnitude (see Fig. \ref{fig_JERM-2020-05-0074_Fig4}.a).
	
	Following the approach described in Section \ref{sec_fitting}, a bi-variate Gaussian mask with $a=1.29\mbox{e}5$ V$^2$/m$^2$ and $\sigma_x=\sigma_y=1.8$ cm has been created to fit the squared amplitude of the E-field corresponding to the SAR distribution reported in Fig. \ref{fig_JERM-2020-05-0074_Fig3}.b (see Fig. \ref{fig_JERM-2020-05-0074_Fig4}.a). 
	
	Computation of the objective function was then performed for SAR center points placed at the nodes of a grid of $N\ped{rfn}=347$ evenly spaced points on the circular refinement region $\mathcal{S}_\mathcal{R}$, with a inter-node spacing $\Delta/6\sim 1.3$ mm. For each grid point, the temperature map was computed solving a 2D version of the bioheat equation \eqref{eq_fII1}, and the corresponding $\tau_{90}$ parameter was evaluated.
	
	\begin{figure*}[!t]
		\centering
		\includegraphics[scale=0.483]{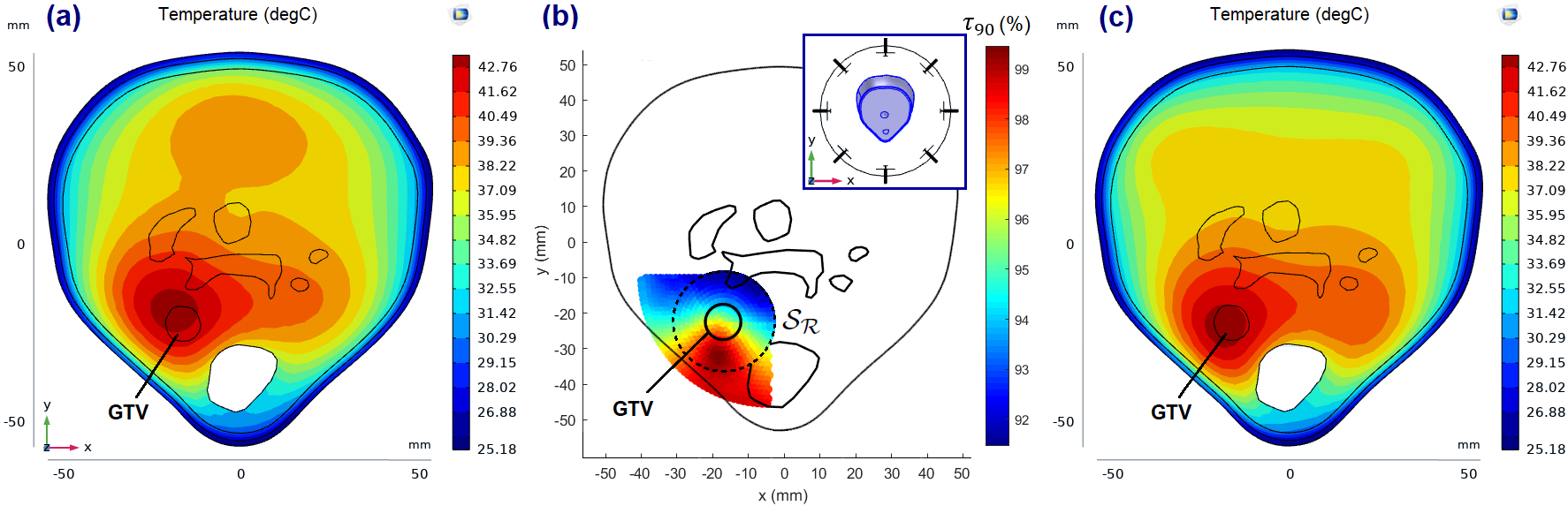}
		\caption{Effect of optimization of SAR target location: realistic neck model. Panel (a): temperature map displayed on the $z=0$ plane, obtained focusing the SAR on the GTV. Panel (b): $\tau_{90}$ parameter as a function of the Gaussian SAR focusing center in the two-dimensional refinement region $\mathcal{S}_\mathcal{R}$. The position corresponding to the maximum $\tau_{90}$ provides the Gaussian SAR focusing center $(\bar{x}_0,\bar{y}_0)$ that optimizes the tumor temperature coverage. Panel (c): temperature distribution on the $z = 0$ plane for a SAR optimized to be focused on a spherical region centered at the pre-corrected location $\bar{\bm{r}}_0=(\bar{x}_0,\bar{y}_0,0)$.}
		\label{fig_JERM-2020-05-0074_Fig6}
	\end{figure*}
	
	Fig. \ref{fig_JERM-2020-05-0074_Fig5}.b shows the $\tau_{90}$ parameter as function of the centers $(x_0,y_0)$ of the Gaussian mask inside the refinement region. The position corresponding to the maximum $\tau_{90}$ provides the Gaussian SAR center $(\bar{x}_0,\bar{y}_0)$ that optimizes the uniformity of the tumor temperature coverage. In the considered case, it was found: $(\bar{x}_0,\bar{y}_0)=(18,-33)$ mm.
	
	The antenna feedings optimization, performed on the 3D model according to \ref{sec_SAR}, maximizing the SAR in a target sphere of diameter $d$ centered around $\bar{\bm{r}}_0=(\bar{x}_0,\bar{y}_0,0)$, provides the temperature map displayed in Fig. \ref{fig_JERM-2020-05-0074_Fig5}.c on the $z=0$ plane. By comparing Fig. \ref{fig_JERM-2020-05-0074_Fig5}.a and Fig. \ref{fig_JERM-2020-05-0074_Fig5}.c, it appears evident how the implemented shift correction has significantly improved the temperature coverage of the GTV. This is quantitatively confirmed by the increase of the $\tau_{90}$ parameter from 95\% to 98\% (Fig. \ref{fig_JERM-2020-05-0074_Fig4}). We then performed a sensitivity study. Using the antenna coefficients derived before and after the application of the T optimization procedure for the tissue baseline parameters of Table \ref{tab_param}, we evaluated the temperature maps when: a) the tumor perfusion is increased by 100\% of its value; b) the perfusion parameters of both tumor and muscle are increased by 100\% of their values; c) the tumor and muscle perfusion parameters are decreased or increased, respectively, by 50\% of their baseline values. In the considered cases, $\tau_{90}$ increases from 96\% to 99\% (a), from 97\% to 99\% (b), and from 96\% to 98\% (c), showing how the improvement is preserved in spite of the perfusion uncertainty.

	\subsection{Realistic neck model} \label{sec_Real}
	The temperature shift correction procedure was then repeated for the more realistic neck computational model previously described in Section \ref{sec_model} and visualized in Fig. \ref{fig_JERM-2020-05-0074_Fig2}. 
	
	As per our proposed method, in the first step, the SAR-based optimization aimed at maximizing the SAR coverage of the tumor was performed. Then, using the same thermal boundary conditions as those reported in Section \ref{sec_BHeq} for the simple model, the corresponding temperature map was obtained (see Fig. \ref{fig_JERM-2020-05-0074_Fig6}.a), showing a shift in the temperature focusing outside the GTV. The corrective strategy described in Section \ref{sec_Correction} was applied again on a two-dimensional refinement region $\mathcal{S}_{\mathcal{R}}$ on the $z=0$ plane, centered at $\bm{r}=\bm{r}_t=(-16.5,-23,0)$ mm, with a diameter $d_{\scriptstyle{\mathcal{R}}}=2.6$ cm and a sampling distance of 1.3 mm, resulting in a total number of points $N\ped{rfn}=347$. 
	
	The search for the optimal Gaussian SAR peak location, i.e. providing the largest value of the objective function $\tau_{90}$, is reported in Fig. \ref{fig_JERM-2020-05-0074_Fig6}.b; the position corresponding to the maximum $\tau_{90}$ was found to be $\bm{\bar{r}}_0=(-18.5,-33.5,0)$ mm.
	
	Finally, the SAR focusing (optimization of Section \ref{sec_SAR}) on a target sphere of diameter $d=10$ mm centered around $\bm{\bar{r}}_0$ led to the temperature map reported in Fig. \ref{fig_JERM-2020-05-0074_Fig6}.c. As before, a significant improvement in the temperature coverage of the GTV on the $z=0$ plane is observed, and confirmed by the increase of the $\tau_{90}$ parameter from 97\% to 99\%.

	\subsection{Numerical Complexity}\label{sec_NumCompl}
With reference to Section \ref{sec_strategy}, the total computational time of our approach is $\timeT^{tot}\ped{TSAR}=\timeT\ped{\:EM}+\timeT\ped{\:TSAR}$, where:
 $\timeT \ped{\:EM}$ is the time for evaluating the antenna fields (step 1),  $\timeT\ped{\:TSAR} = \timeT\ped{\:search} +2 \timeT\ped{\:SAR}$ the time for steps 2, 5 and 7, with  $\timeT\ped{search}$ the cost of the optimum SAR focusing search, and $\timeT \ped{\:SAR}$ the time of a SAR optimization. The latter is $\timeT\ped{\:SAR}=N\ped{OPT}\cdot  t\ped{LC}$, with $N\ped{OPT}$ the total number of iterations of the SAR optimization algorithm, and $t\ped{LC}$ the time to compute a superposition of the antenna fields. For $N\ped{rfn}$ points in the refinement process,  $\timeT\ped{search}=N\ped{rfn}\cdot t\ped{bioH}$, with $t\ped{bioH}$ the time needed to solve the bioheat equation.
	
To estimate the cost of a direct temperature optimization, $\timeT^{tot}\ped{T}$, we consider a baseline procedure with a global optimization requiring $N\ap{T}\ped{OPT}$ steps; each step involves a solution of the bioheat equation, while the SAR is obtained via superposition of the individual antenna fields, i.e., $	\timeT^{tot}\ped{T}=\timeT\ped{\:EM}+\timeT\ped{\:T}$, $\timeT\ped{\:T} =N\ap{T}\ped{OPT} \cdot (t\ped{LC} + t\ped{bioH})$. 

In our example (realistic H\&N), the PSO required $N\ped{OPT}=20\mbox{e}3$ iterations. The time needed to compute a combination of the antenna fields was $t\ped{LC}\approx 0.006$ s, on the considered workstation and with our non-optimized MATLAB code. In the refinement process, $N\ped{rfn}=347$ and a 2D solution of the bioheat equation was always found to be enough, requiring a computational time $t\ped{bioH,2D}\approx 1.2$ s much lower than the time needed for the 3D case ($t\ped{bioH,3D}\approx 10$ s). This results in $\timeT\ped{search}\approx 7$ min, $\timeT\ped{\:SAR}\approx 2$ min, and hence $\timeT\ped{\:TSAR}\approx 11$ min.

It is reasonable to assume that (as here) $ t\ped{LC} \ll t\ped{bioH}$ in general, and that $N\ap{T}\ped{OPT} \sim N\ped{OPT}\gg N\ped{rfn}$, which leads to the following estimates:
$
\timeT^{tot}\ped{\:T} - \timeT^{tot}\ped{\:TSAR} \sim \left(N\ped{OPT}-N\ped{rfn}\right) \cdot t\ped{bioH}-N\ped{OPT}\cdot t\ped{LC} \sim  N\ped{OPT} \cdot (t\ped{bioH}-t\ped{LC})=\timeT\ped{\:SAR}\cdot\left(\frac{t\ped{bioH}-t\ped{LC}}{t\ped{LC}}\right) \sim \timeT\ped{\:SAR}\cdot\frac{t\ped{bioH}}{t\ped{LC}}.
$

	\section{Conclusion}
	We proposed a temperature optimization via a SAR optimization, but with an optimized target. Hence, it can also be added to existing SAR-based procedures with minor modifications. The light computational requirements, and its SAR-based nature also allows sensitivity studies with respect to critical thermal parameters, like perfusion, that are known with significant uncertainties.
	
	The low complexity is afforded by optimizing the position of the center of the SAR target; stronger distortions of the T map with respect to SAR map are not accounted for, and this may constitute a limitation of the present method. 
	
	Future work will be primarily devoted to testing the method in other scenarios with more accurate anatomical models and to improve the SAR target optimization. We are currently analyzing the possibility to implement 3D search and reduced-order modeling approaches \cite{DEIM1,DEIM}. Convex optimization is being investigated for target optimization. Furthermore, the current Gaussian approximation can be removed and substituted by multiple SAR optimizations.

	\section*{Acknowledgment}
	This work has been supported by the Italian Ministry of Research under PRIN ``Field and Temperature Shaping and Monitoring for Microwave Hyperthermia FAT SAMMY''.

	\begin{IEEEbiography}[{\includegraphics[width=1in,height=1.25in,clip,keepaspectratio]{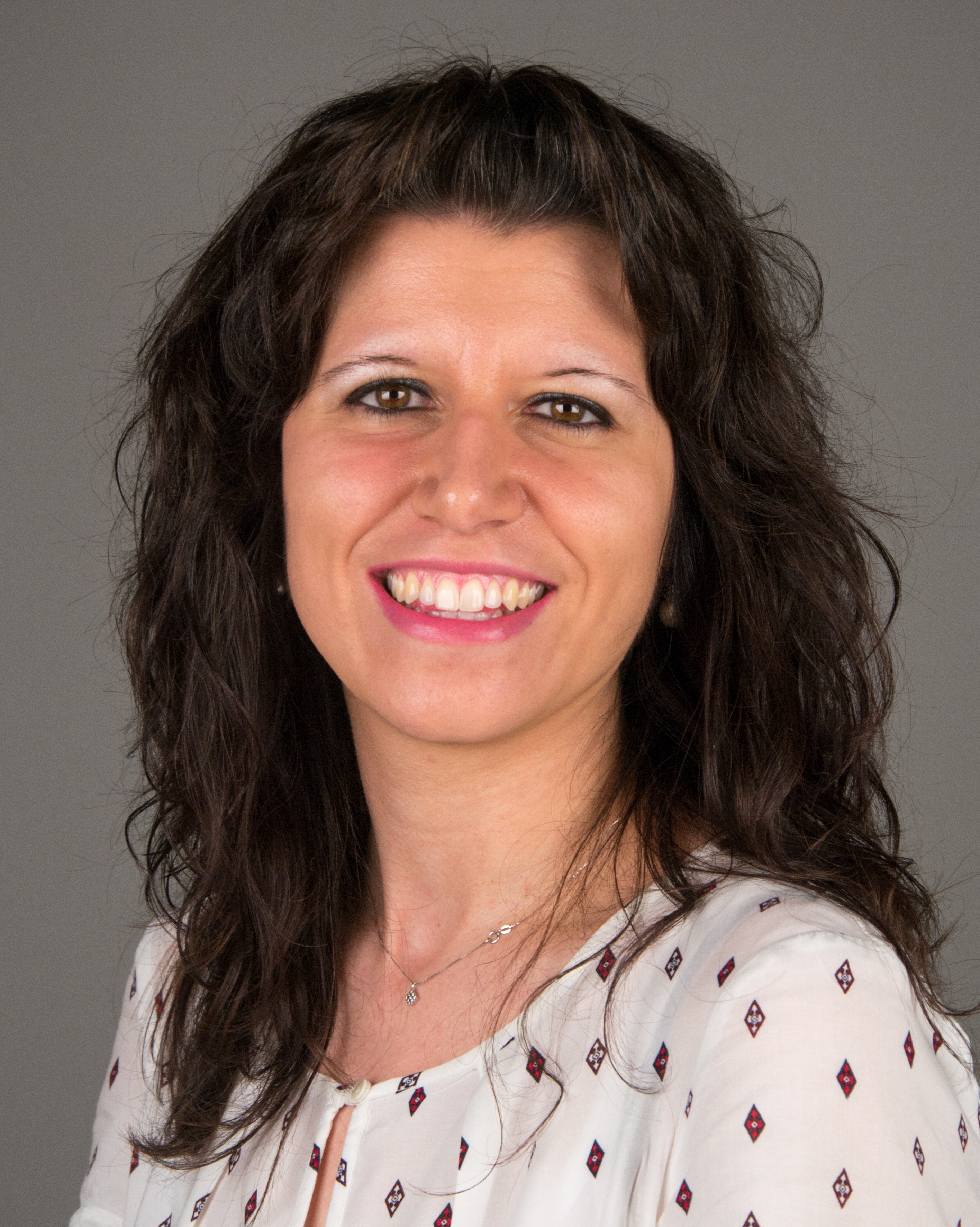}}]{Rossella Gaffoglio}
		received the M.Sc. degree (cum laude) in physics and the Ph.D. in physics and astrophysics from the University of Torino, Italy, in 2013 and 2017, respectively. From 2014 to 2016, during the Ph.D. period, and in 2017, as a scholarship holder of the University of Torino, she collaborated as a consultant with the Centre for Research and Technological Innovation, RAI Radiotelevisione Italiana, focusing her research activity on the analysis of electromagnetic waves with unusual topologies, array synthesis, multiplexing/demultiplexing schemes, numerical algorithms for mobile and fixed TV networks. In October 2017 she joined the Department of Electronics and Telecommunications, Politecnico di Torino, as a Research Associate, where she worked on the project: ``Field and Temperature Shaping and Monitoring for Microwave Hyperthermia''. Since June 2018 she is a researcher at the Advanced Computing and Applications (ACA) area of Fondazione LINKS, where her research activity mainly concerns the analysis and numerical modelling of antenna systems, medical applications of electromagnetic fields and imaging.
	\end{IEEEbiography}
	
	\begin{IEEEbiography}[{\includegraphics[width=1in,height=1.25in,clip,keepaspectratio]{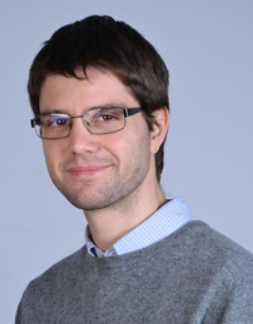}}]{Marco Righero}
		received the B.Sc. degree in Mathematics for Engineering Sciences in 2003, the M.Sc. degree (cum laude) in Mathematical Engineering in 2005, and the European Ph.D. degree in Electronic and Communication Engineering in 2009, all from the Politecnico di Torino, Italy. His main research activity, carried out at the Politecnico di Torino, the University of California San Diego (U.S.A.), and the University College Cork (Ireland), focused on synchronization phenomena in complex networks and biologically plausible circuit models of neurons. In 2009 he joined the Electronics Department of Politecnico di Torino as a research assistant with a fellowship from the Istituto Superiore Mario Boella (ISMB, now part of Fondazione LINKS), working on nonlinear system and signal processing algorithms for bio-inspired sensing devices. Within this project, he received a grant from the Fondazione CRT. In 2010 he started collaborating with the Antenna and EMC Lab (LACE) of ISMB and in 2011 he officially joined the group, where he works on innovative models and efficient methods for the study of electromagnetic propagation in complex media, optimization techniques for antenna arrays synthesis, and antenna measurements.
	\end{IEEEbiography}
	\vfill
	
	\begin{IEEEbiography}[{\includegraphics[width=1in,height=1.25in,clip,keepaspectratio]{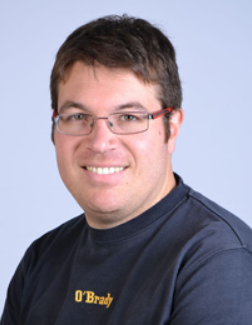}}]{Giorgio Giordanengo}
		received the B.Sc. degree in electronics engineering, the M.Sc. degree in telecommunications engineering, and the Ph.D. degree in the scientific area of information and communications engineering (with a special focus on applied electromagnetics) from the Politecnico di Torino, Turin, Italy, in 2006, 2009, and 2016, respectively. In 2010, he obtained the Engineering Professional Qualification in the information field with the Politecnico di Torino. In 2008, as part of his M.Sc. thesis, he was with the PLAN Group, University of Calgary, Calgary, AB, Canada, where he was involved in the field of availability of GNSS services also in the presence of RF interference, like jammers. From 2009 to 2011, he was a System Design Engineer with Alenia SIA SpA, Turin, where he was involved with research on the design, development, and test for new functionalities to add into the Eurofighter Typhoon navigation subsystem. In 2011, he joined the Antenna and EMC Lab, Istituto Superiore Mario Boella, Turin. He is currently a Researcher with Fondazione LINKS, Turin. His research interests include innovative algorithms able to test large structures at the RF level by mixing simulations and measurements, design and development of innovative antennas up to Ka-band, and exploiting the latest antenna technologies like liquid crystals and antenna testing and measurements (both in an anechoic chamber rather than in outdoor).
	\end{IEEEbiography}
	\vfill
	
	\begin{IEEEbiography}[{\includegraphics[width=1in,height=1.25in,clip,keepaspectratio]{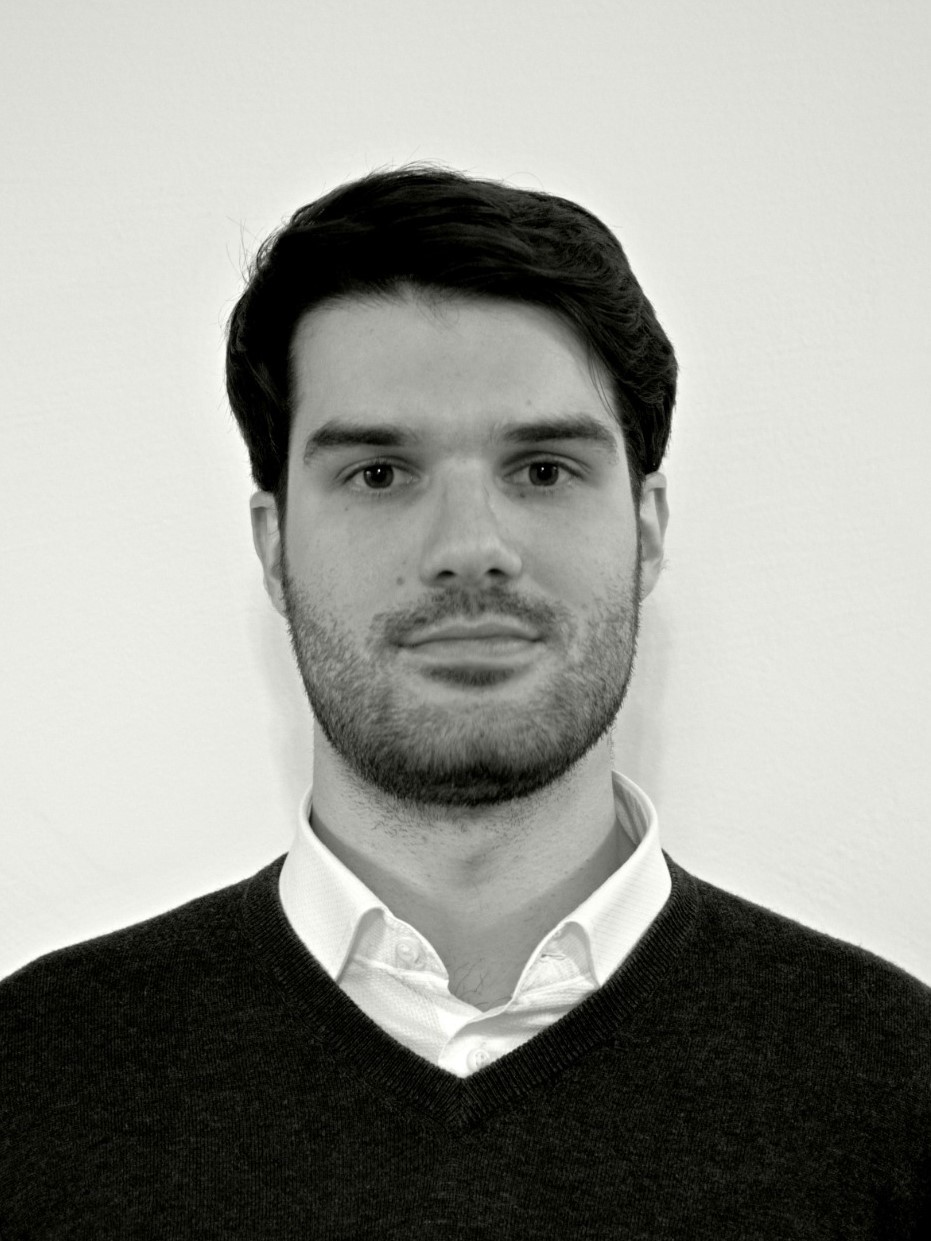}}]%
		{Marcello Zucchi}
		received the B.Sc. degree in Electronics and Telecommunications Engineering from the University of Bologna, Bologna, Italy and the M.Sc. degree in Electronics Engineering from Politecnico di Torino, Turin, Italy, in 2014 and 2018, respectively. He is currently pursuing his Ph.D. in Electrical, Electronics and Communications Engineering at the Politecnico di Torino. His research interests include global optimization algorithms for flat antenna design, automatic synthesis of modulated metasurface antennas and field focusing for hyperthermia treatment.
	\end{IEEEbiography}
		
	\begin{IEEEbiography}[{\includegraphics[width=1in,height=1.25in,clip,keepaspectratio]{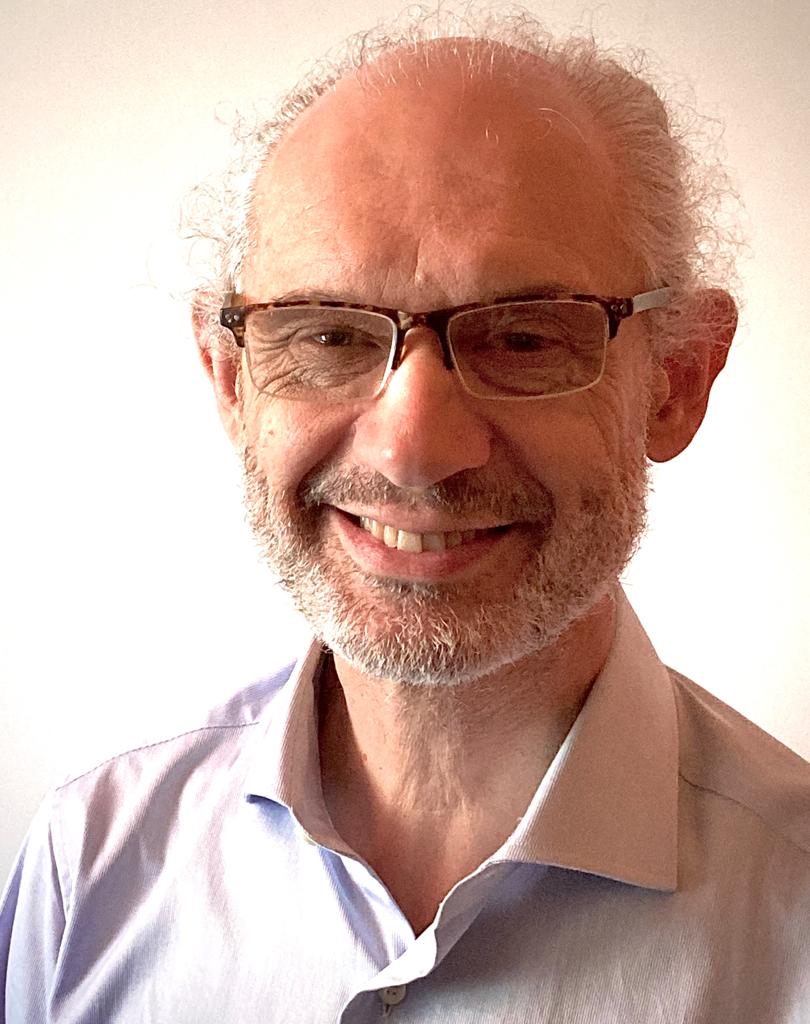}}]%
		{Giuseppe Vecchi}
		(Fellow, IEEE) received the Laurea and Ph.D. degrees in electronic engineering from the Politecnico di Torino, Turin, Italy, in 1985 and 1989, respectively, with doctoral research carried out partly at Polytechnic University, Farmingdale, NY, USA. He was a Visiting Scientist with Polytechnic University from 1989 to 1990. Since 1990, he has been with the Department of Electronics, Politecnico di Torino, as an Assistant Professor, an Associate Professor in 1992, and a Professor since 2000, where he is currently the Director of the Antenna and Electromagnetic Compatibility Laboratory. He was also a Visiting Scientist with the University of Helsinki, Helsinki, Finland, in 1992 and an Adjunct Faculty Member with the Department of Electrical and Computer Engineering, University of Illinois at Chicago, Chicago, IL, USA, from 1997 to 2011. His current research activities concern analytical and numerical techniques for analysis, design, and diagnostics of antennas and devices, and imaging. Prof. Vecchi is a member of the Board of the European School of Antennas; he has been the Chairman of the IEEE AP/MTT/ED Italian joint Chapter on the IEEE-APS Educational Committee and an Associate Editor of the IEEE Transactions on Antennas and Propagation.
	\end{IEEEbiography}

\end{document}